\documentclass{elsart}

\usepackage{natbib}

\usepackage{amssymb}

\begin{document}

\begin{frontmatter}




\title{Lessons Learned from the Pioneers 10/11 for \\
A Mission to Test the Pioneer Anomaly}


\author[label1]{Slava G. Turyshev\corauthref{cor1}},
\corauth[cor1]{Corresponding author.}
\ead{turyshev@jpl.nasa.gov}
\author[label2]{Michael Martin Nieto},
\author[label1]{John D. Anderson}
\address[label1]{Jet Propulsion Laboratory,
California Institute of Technology,\\
4800 Oak Grove Drive, Pasadena, CA 91109}

\address[label2]{Los Alamos National Laboratory,
University of California, \\Los Alamos, NM 87545}

\begin{abstract}

Analysis of the radio-metric tracking data from the Pioneer 10/11
spacecraft at distances between 20--70 astronomical units (AU) from
the Sun has consistently indicated the presence of an anomalous,
small, constant Doppler frequency drift.  The drift is a blue-shift,
uniformly changing with rate $a_t = (2.92 \pm 0.44)\times 10^{-18}$
s/s$^2$.  It can also be interpreted as a constant acceleration of
$a_P  = (8.74 \pm 1.33) \times  10^{-8}$ cm/s$^2$ directed towards the
Sun.  Although it is suspected that there is a systematic origin to
the effect, none has been found.  As a result, the nature of this
anomaly has become of growing interest.  Here we discuss the details
of our recent investigation focusing on the effects both external to and
internal to the spacecraft, as well as those due to modeling and
computational techniques.  We review some of the mechanisms proposed
to explain the anomaly and show their inability to account for the
observed behavior of the anomaly.  We also present lessons learned
from this investigation for a potential deep-space experiment that
will reveal the origin of the discovered anomaly and also will
characterize its properties with an accuracy of at least two orders of magnitude below the anomaly's size.  A number of critical requirements and design considerations for such a mission are outlined and addressed.  
\end{abstract}


\end{frontmatter}

\section{The Pioneer Missions and the Anomaly}
\label{sec:background}

The Pioneer 10/11 missions, launched on 2 March 1972 (Pioneer 10) and 4 December 1973 (Pioneer 11), were the first to explore the outer solar system.  After Jupiter and (for Pioneer 11) Saturn encounters, the two spacecraft followed escape hyperbolic orbits near the plane of the ecliptic to opposite sides of the solar system.  Pioneer 10 eventually became the first man-made object to leave the solar system.  
 
The Pioneers were excellent craft with which to perform precise celestial mechanics experiments.  This was due to a combination of many factors, including their attitude control (spin-stabilized, with a minimum number of commanded attitude correction maneuvers using thrusters), power design (the RTGs being on extended booms aided the stability of the craft and also reduced the heat systematics), and precise Doppler tracking (with sensitivity to resolve small frequency drifts at the level of mHz/s).  The result was the most precise navigation in deep space to date. 

By 1980 Pioneer 10 had passed a distance of $\sim$20 AU from the Sun and the acceleration contribution from solar-radiation pressure on the craft (directed away from the Sun) had decreased to less than $4 \times 10^{- 8}$ cm/s$^2$.  At that time an anomaly in the Doppler signal became evident.  Subsequent analysis of the radio-metric tracking data from the Pioneer 10/11 spacecraft at distances between 20--70 AU from the Sun has consistently indicated the presence of an anomalous, small, constant Doppler frequency drift.  The drift can be interpreted as being due to a constant acceleration of $a_P  = (8.74 \pm 1.33)\times 10^{-8}$ cm/s$^2$ directed towards the Sun \citep{prl_98,prd_01,mpla_01}. The nature of this anomaly remains unexplained; that is to say, up to now no satisfactory explanation of the anomalous signal has been found.  This signal has become known as the Pioneer Anomaly.

Although the most obvious explanation would be that there is a systematic origin to the effect, perhaps generated by the spacecraft themselves from excessive heat or propulsion gas leaks, none has been found; that is, no unambiguous, onboard systematic has been discovered \citep{prl_98,prd_01,Markwardt_02}.  In fact, attempts to find a convincing explanation using such a mechanism have not succeeded.  This inability to explain the anomalous acceleration of the Pioneer spacecraft with conventional physics has contributed to the growing discussion about its origin. 

Attempts to verify the anomaly using other spacecraft have proven disappointing \citep{ijmpd_02,Nieto_Turyshev_cqg_2004}.  This is because the Voyager, Galileo, Ulysses, and Cassini spacecraft navigation data all have their own individual difficulties for use in an independent test of the anomaly.  In addition, many of the deep space missions that are currently being planned either will not provide the needed navigational accuracy and trajectory stability of under $10^{-8}$ cm/s$^2$ (i.e., Pluto Express, Interstellar Probe) or else they will have significant on-board systematics that mask the anomaly (i.e., JIMO -- Jupiter Icy Moons Orbiter). 

The acceleration regime in which the anomaly was observed diminishes the value of using modern disturbance compensation systems for a test.  For example, the systems that are currently being developed for the LISA and LISA Pathfinder missions, are designed to operate in the presence of a very low frequency acceleration noise (at the mHz level), while the Pioneer anomalous acceleration is a strong constant bias in the Doppler frequency data.  In addition, currently available DC accelerometers are a few orders of magnitude less sensitive than is need for a test.  Furthermore, should the anomaly be a fictitious force that universally affects frequency standards \citep{prd_01}, the use of accelerometers will shed no light on what is the true nature of the observed anomaly. 

Finally, a comprehensive test of the anomaly requires an escape hyperbolic trajectory \citep{prd_01,ijmpd_02,Nieto_Turyshev_cqg_2004}.  This makes a number of advanced missions (i.e., LISA -- the Laser Interferometric Space Antenna, STEP -- the Space Test of Equivalence Principle, LISA Pathfinder) less able to test properties of the detected anomalous acceleration.  Although these missions all have excellent scientific goals and technologies, nevertheless, because of their orbits they will be in a less advantageous position to conduct a precise test of the detected anomaly. 

The inability to find a standard explanation for the anomaly, combined with the evident lack of suitable experimental opportunities, motivated an interest in developing a designated mission to study the detected signal. 
%
%
Here we focus on the lessons learned from our previous Pioneer anomaly investigation, especially on their relevance for a potential new deep-space experiment.  The mission could lead to a determination of the origin of the discovered anomaly; it could also characterize its properties to an accuracy of at least three orders of magnitude below the anomaly's size.  The mission must be capable to discover if the anomaly is due to some unknown physics or else to an on-board systematic.  Either way the result would be of major significance.  If the anomaly is a manifestation of new or unexpected physics, the result would be of truly fundamental importance.  However, even if the anomaly turns out to be an unknown manifestation of an on-board systematic, its understanding would vitally affect the design of future precision space navigation, especially in deep space.  Furthermore, technologies and mission design solutions envisioned for this experiment will be vital to many space missions that are to come. 

This paper is organized as follows: in Section 2 we discus the Pioneer anomalous acceleration and attempts to explain it.  In Section 3 we will discuss the mission objectives for this proposed mission. This section highlights the lessons learned from the Pioneers 10/11, as they apply to a designated mission to test the Pioneer anomaly.  We close with a summary and recommendations.

\section{Attempted Explanations of the Pioneer Anomaly}
\label{sec:explanations}

After the announcement of the anomalous acceleration, many proposals appeared that invoked conventional physics to explain its origin.  To do this one needs to find a systematic origin.  However, up to now no satisfactory explanation of this type has been found.  This was summarized in \citep{prd_01}, where possible contributions of various mechanisms to the final error budget of the solution for the anomalous acceleration were given.  The error budget was subdivided into three main types of effects that could contribute to the anomaly.  Below we present a summary of the conclusions:  

The first group of effects were those external to the spacecraft; such as the solar radiation pressure, effects of the solar wind, and the effect of the solar corona on the propagation of radio-wave signals.  In addition, Ref. \citep{prd_01} discussed the influence of the Kuiper Belt's and the galaxy's gravity, the influence of the interplanetary dust in the solar system, electro-magnetic Lorentz forces, and errors in the accepted values of the Earth's orientation parameters, precession, and nutation.  The analysis evaluated the contributions of the mechanical instabilities and the location errors of the DSN antennae structures, the phase stabilities of the DSN antennae and clocks, and effects due to the troposphere and ionosphere.  None of these effects even came close to providing an expiation of the anomaly.  Even though some of these mechanisms are near the limit for contributing to the final error, it was found that none of them could explain the found signal, and some were three orders of magnitude or more too small.  In totality, they were insignificant.

The second group of effects were those that originated on-board and are tied to a well-known sources; this group, as expected, had the largest impact to the final error.  Among these effects, the radio beam reaction force produced the largest bias to the result, $1.10 \times  10^{-8}$ cm/s$^2$.  It actually made the Pioneer effect larger.  Large uncertainties also came from differential emissivity of the Radioisotope Thermoelectric Generators (RTG), radiative cooling, and propulsive gas leaks from thrusters of the attitude control system, $\pm 0.85, \pm0.48$, and  $\pm0.56$, respectively,  $10^{-8}$ cm/s$^2$.  The effect due to expelled Helium produced within the RTGs was also considered, so as the small difference in anomaly determinations between the two Pioneers.

But it is he second largest bias/uncertainty, from the on-board heat rejected from the spacecraft, that has been the most critical systematic to quantify.  As is known \citep{prl_98,prd_01,mpla_01}, the Pioneer spacecraft were powered with the SNAP-19 RTGs mounted on long extended booms (designed to protect the on-board electronics from heat and radiation impact).  In, principle, there was more than enough heat available on the craft to cause the anomaly.  However, the spacecraft spin-stabilized attitude control, the special design for the RTGs used on the Pioneers, and the length of the RTG booms, which resulted in a relatively small spacecraft surface available for preferential heat reflection, significantly minimized the amount of heat for the mechanism to work.  The analysis of the 11.5 years of Pioneer Doppler data \citep{prd_01} can only support effect as large as $(-0.55\pm0.55) \times 10^{-8}$ cm/s$^2$.  

In summary, this second group represented the most likely sources for the anomaly.  However, none of these mechanisms gained enough experimental support to explain the anomaly.  At most one can obtain $\sim$12\% of the discovered effect by employing all of these mechanisms.  Furthermore, there was no obvious ``smoking gun'' found in this category of effects.   

The third group of effects were composed of contributions from computational errors.  The effects in this group dealt with numerical stability of least-squares estimations, accuracy of consistency/model tests, mismodeling of maneuvers, and the solar corona model used to describe the propagation of radio waves.  \citet{prd_01} also analyzed the influence of annual/diurnal terms seen in the data on the accuracy of the estimates.  These effects were all small.

These three groups of effects exhausted all available conventional explanations for the anomaly.  The inability to explain the Pioneer anomaly with conventional physics has led to a significant number of theoretical proposals that use more unusual mechanisms (more details are in \citep{prd_01}).  As time progresses, the number of new ideas is increasing and some of these have strong science potential and warrant a new experimental investigation.  Thus, the Pioneer anomaly made our own backyard - the solar system - a new terra incognita.  

In conclusion, there are two main possible explanations of the origins for the detected anomalous acceleration.  The first is on-board generated systematics.  Dispassionately, this is the most likely cause of the anomaly, but until now the smoking gun still had not been found.  The second possible origin is new physics.  As noted previously, this dichotomy represents a healthy 'win-win' situation; either one of these two possibilities would be an extremely important discovery.  If the anomaly is due to some not-yet understood systematic, our understanding of it would help us to build more stable and less noisy spacecraft that can be navigated more precisely for the benefit of deep-space fundamental physics experiments in the 21st century.  If the anomaly is due to new physics, the possible implications of this opportunity are enormous.

\section{A Mission to Test the Pioneer Anomaly}
\label{sec:mission}

In this section we address the possibility of accomplishing the test by using existing spacecraft technologies in combination with newly developed capabilities.  We argue that such a mission could be an excellent opportunity to develop and demonstrate new technologies for spacecraft design, in-space propulsion, on-board power, and others that will find their way into many other applications for space exploration in the 21st century.

\subsection{Mission Objectives}
\label{sec:mission_obj}

The main scientific goal of this deep space mission is to determine the origin of the Pioneer anomalous Doppler frequency drift and to characterize its properties to an accuracy of $\sim0.01 \times 10^{-8}$ cm/s$^2$; that is, two to three orders of magnitude below the inferred anomaly's size.   Similarly, the investigation of possible clock accelerations is proposed to be carried out with the sensitivity of $\sim 3 \times 10^{-21}$ s/s$^2$.  The scientific merit of this mission has already been discussed \citep{ijmpd_02,Nieto_Turyshev_cqg_2004}.  We emphasize its possible importance in unraveling unknown physics and also its ability to provide a significant accuracy improvement in the methodology for characterizing small anomalous accelerations.

As far as the technologies are concerned, their development might turn out to be the most important result of the proposed mission.  Future precision experiments in very deep space are currently being envisioned.  As is vividly demonstrated by the Pioneers, the effects of small systematic forces are not easily modeled and compensated, even today. Further, communication frequency drifts are generally not monitored at the required level and, in fact, these levels are a few orders below the desired sensitivity. Therefore, understanding the anomaly in terms of our precisely conceived mission craft would help engineers design and build more stable and less noisy spacecraft in the future. 

\subsection{Applying lessons learned from the Pioneers}
\label{sec:applying_lessons}

The lessons learned from the Pioneers are a guide on how to build a spacecraft that will achieve our goals. Among the most important features of the Pioneers were their attitude control system, navigation and communication, on-board power, thermal design, and mission design \citep{Nieto_Turyshev_cqg_2004}.

These lessons allow us to suggest major features that are needed on a possible mission to test the Pioneer Anomaly, especially those related to the mission's spin-stabilization, on-board power, and ``fore/aft'' symmetric bus and antenna designs.  They also give input into the hyperbolic escape orbit (and launch concept of the next section). Below we present a summary of our findings on the features that are critical for our proposed mission. 

\subsubsection{Attitude control} 
The ultimate goal for the attitude control system is to enable a 3D acceleration sensitivity to the level of $\sim0.01 \times 10^{-8}$ cm/s$^2$ for each spacecraft axis.  As with the Pioneers, this can be done with spin-stabilized attitude control, which is preferred for our mission.  This choice allows for a minimum number of attitude correction maneuvers which are, because of the maneuver-associated propulsive gas leaks, notoriously difficult to model.  Leakage from thrusters of the propulsion system is the major navigation problem for 3-axis stabilized vehicles, but its impact is minimal for spin-stabilized spacecraft.   If spin-stabilization is chosen, spacecraft spin behavior can be precisely monitored.  The understanding of the spin history, coupled with knowledge of all possible sources of torque, will provide auxiliary information on the anomaly. 

\subsubsection{On-board propulsion system} 
For the reasons discussed in the attitude control requirements above, one would need precisely calibrated thrusters, propellant lines, fuel gauges and knowledge of the propellant mass usage history.  However, currently available sensors are not sensitive enough for our purposes.   Since their information may be critical for the precise orbit solutions we desire, we strongly suggest further development of these technologies. Autonomous real-time monitoring and control of their performances would also be a big plus.  

\subsubsection{Navigation and communication} 
As with the attitude control system, the navigation and communication system should allow a 3D acceleration reconstruction at the level of $\sim 0.01 \times 10^{-8}$ cm/s$^2$ for each vector component.  Having both Doppler and range tracking, and possibly VLBI and/or $\Delta$DOR, will allow the precise measurements of plane-of-the-sky angles that are needed for 3D acceleration reconstruction.  A $\mu$rad pointing should be sufficient to enable precise attitude reconstruction.  The preferred communications frequencies are X- and Ka-band with significant dual-band tracking.  (Once optical tracking has been successfully demonstrated in space, it will be a desirable alternative.)  

\subsubsection{On-board power} 
No deep-space mission can accomplish its goals without a reliable source of on-board power.  For now this must be provided by RTGs. Location of the RTGs is a very critical choice, as it must provide inertial balance, stability and thermal isolation.  For a spin-stabilized option, one would want to position the RTGs as far as practical from the bus. Having the RTGs on extended booms aids the stability of the craft and also reduces the heat systematics.  If 3-axis stabilization were employed, the effects of the frequent gas-jetting would not allow the navigational precision we need unless accelerometers and reaction wheels were used.

\subsubsection{Thermal design} 
This is one of the most critical designs for our mission, as the emitted radiant heat from the RTGs must be symmetrical in the fore and aft directions.  For a spin-stabilized craft, the thermal louvers will be placed on the sides to eliminate the thermal recoil force due to release excess of radiant heat.   Furthermore, the entire spacecraft should be heat-balanced and heat-symmetric.  One would also need to have precise knowledge of all heat sources - RTGs, electronics, thrusters, etc.  In addition, an active control of all heat dissipation channels is also a critical requirement.  Finally, it is also important to have a precise knowledge of degradation the spectral properties of materials from which the spacecraft surface is composed. This all should result in the precise knowledge of the history of the 3D vector of a residual thermal recoil force, if any.

\subsubsection{Symmetric radio-beam}  
A for/aft symmetric spacecraft design uses two identical and simultaneously transmitting radio-antennae placed facing opposite directions along the spin-axis.   By implementing such a design, one significantly reduces the radio-beam communication bias and also the preferential thermal recoil-force-induced acceleration bias \citep{Nieto_Turyshev_cqg_2004}.  This choice would also eliminate any remaining fore/aft asymmetry in the acceleration estimation by periodically rotating the craft by 180$^\circ$.  

\subsubsection{Hyperbolic, solar system escape orbits}  
The Pioneer anomaly was found on craft following hyperbolic, un-bound, escape trajectories at distances between 20 and 70 AU out from the Sun \citep{prd_01}.  Although, it might have been present closer in, this has only been imprecisely studied \citep{prl_98,prd_01,ijmpd_02,Nieto_Turyshev_cqg_2004}. For this reason and also to reduce the effect of external systematics the experiment should reach distances greater than 15 AU from the Sun.  Obviously, one wants a fast orbit transfer to this region; say, not much more than 6 years.  To yield a direct test for any velocity-dependence in the signal, one also wants the craft to have a significantly different velocity than the Pioneers.  All this means that when the craft reaches deep space it should be in a high-velocity, hyperbolic, escape orbit.  

Concluding, the Pioneers were ``accidentally'' built in a way that yielded very precise orbits; newer craft will need special designs to surpass their accuracies.  Effects that have normally been considered to be relatively unimportant, rejected thermal radiation, gas leaks, and radio beam reaction, now turn out to be critical for the precise navigation of science craft in the 21st century.  It is hard not to emphasize the most successful feature and main Pioneer lesson for a potential spacecraft and mission design to test the anomalous acceleration - make it simple!  

\subsection{Propulsion options}
\label{sec:propulsion}

The launch vehicle is a major consideration for any deep-space mission. To test the Pioneer anomalous acceleration in the most suitable environment, one wants to reach a distance greater than 15 AU from the Sun.   In this region one can clearly distinguish any effect from solar radiation pressure, interplanetary magnetic fields, as well as solar and interplanetary plasmas.  A fast transfer orbit is very desirable, to allow reaching the target region in a minimal time.  Therefore, a large solar system escape velocity is desired (say, more than 5-10 AU/yr).  In contrast, the Pioneers are cruising at a velocity of about 2 AU/year and the Voyagers at about 3 AU/year. One needs something faster than that.   

Propulsion systems are quite literally the driving force behind any effort to get a payload into space, especially on an interplanetary orbit. Over the years, advances in engine technology have helped chemical propulsion realize significant gains in performance and cost. However, the use of chemical propulsion is at the limit of its capabilities to satisfy the needs for deep-space exploration.  For this reason, both ESA and NASA have initiated programs to study alternative propulsion methods for their deep space exploration missions. 

The obvious first idea is a very energetic rocket with chemical propulsion.  An escape terminal velocity of $\sim$5 AU/yr is achievable with current launch and mission design technologies. It can be done with existing heavy launch vehicles (Ariane V, Proton, Delta IV or Atlas V), \citep{Nieto_Turyshev_cqg_2004,aero_2004}.  Pioneer data taken before escape velocity was reached and starting before the flybys of any major planets was never thoroughly analyzed.  In particular, the Pioneer 11 data roughly indicates that the anomaly started near its Saturn flyby, when it reached escape velocity.   If a chemical launch vehicle were to be used for a dedicated mission, with gravity assist flybys, our mission could also address this question. 

If one were to use chemical propulsion for a major mission with a Pioneer probe to be separated after main launch, this choice would require a very long cruse phase in the inner solar system with multiple planetary fly-bys, before the craft reaches escape trajectory.  In the case of the Cassini spacecraft, it took almost 7 years to reach Saturn (9.5 AU), which is prohibitively long for our mission. We need something faster then that, which makes alternative launch scenarios of even more interest. Thus alternative propulsion concepts may be considered, such as solar-sail or nuclear-electric propulsion.  These options will be further investigated.

\section*{Conclusions}
\label{sec:conclusions}

We presented lessons learned from our recent investigation of the
Pioneer anomalous acceleration. These lessons are important in
studying a design for a potential deep-space experiment that will
reveal the origin of the discovered anomaly and also will characterize
its properties to an accuracy of at least two orders of magnitude
below the anomaly's size.  A potential mission design should be able
to eliminate or significantly minimize the effects of small forces
either external to and internal to the spacecraft, as well as those
due to modeling and computational techniques.  A number of critical
requirements and design considerations for the mission are outlined
and addressed. With the evident interest of the international scientific community in  this investigation, a mission to test the Pioneer anomaly is rapidly approaching a realistic design phase.  Our analysis also indicates that if only existing technologies are used, it
could be developed in about 5 years and flown early in the next
decade. 

The work described here by SGT and JDA was carried out at the Jet Propulsion Laboratory, California Institute of Technology under a contract with the National Aeronautics and Space Administration. MMN acknowledges support by the U.S. Department of Energy.




\end{document}